\documentclass[aip,amsmath,amssymb,
 reprint,twocolumn, superscriptaddress
]{revtex4}

\usepackage{graphicx}
\usepackage{dcolumn}
\usepackage{bm}
\usepackage[utf8]{inputenc}
\usepackage[T1]{fontenc}
\usepackage{mathptmx}
\usepackage[normalem]{ulem} 	
\usepackage[usenames,dvipsnames]{color}
\usepackage{xcolor} 
\usepackage{SIunits} 

\newif\iffinal
\finaltrue

\newcommand{\del}[1]{\sloppy{\textcolor{blue}{\sout{#1}}}} 

\newcommand{\macom}[1]{{\marginpar{\textcolor{blue}{#1}}}} 

\newcommand{\jl}[1]{\textcolor{blue}{#1}}
\newcommand{\ab}[1]{\textcolor{orange}{#1}}

\iffinal
	
	\renewcommand{\del}[1]{}
	\renewcommand{\jl}[1]{}
	\renewcommand{\ab}[1]{}

	\renewcommand{\macom}[1]{}
	
\else
\fi

\begin{document}

\title{In-situ bandaged Josephson junctions for superconducting quantum processors}

\author{Alexander Bilmes}
\email[]{e-Mail: abilmes@google.com}
\affiliation{Physikalisches Institut, Karlsruhe Institute of Technology, 76131 Karlsruhe, Germany}
\author{Alexander K. H\"andel}
\affiliation{Physikalisches Institut, Karlsruhe Institute of Technology, 76131 Karlsruhe, Germany}
\author{Serhii Volosheniuk}
\affiliation{Kavli Institute of Nanoscience, Delft University of Technology, 2628 CJ Delft, Netherlands}
\author{Alexey V. Ustinov}
\affiliation{Physikalisches Institut, Karlsruhe Institute of Technology, 76131 Karlsruhe, Germany}
\affiliation{National University of Science and Technology MISIS, Moscow 119049, Russia}
\affiliation{Russian Quantum Center, Skolkovo, Moscow 143025, Russia}
\author{J\"urgen Lisenfeld}
\affiliation{Physikalisches Institut, Karlsruhe Institute of Technology, 76131 Karlsruhe, Germany}

\date{\today}

\begin{abstract}
	\centering\begin{minipage}{\linewidth}
		\textbf{						
		Shadow evaporation is commonly used to micro-fabricate the key element of superconducting qubits - the Josephson junction. However, in conventional two-angle deposition circuit topology, unwanted stray Josephson junctions are created which contribute to dielectric loss. So far, this could be avoided by shorting the stray junctions with a so-called bandage layer deposited in an additional lithography step, which may further contaminate the chip surface. Here, we present an improved shadow evaporation technique allowing one to fabricate sub-micrometer-sized Josephson junctions together with bandage layers in a single lithography step. We also show that junction aging is significantly reduced when junction electrodes are oxidized in an oxygen atmosphere directly after deposition.	
		}
	\end{minipage}
\end{abstract}

\maketitle 
\setlength{\parskip}{-0.25cm}


In superconducting quantum processors, qubits are realized with non-linear resonators formed by capacitively or inductively shunted Josephson tunnel junctions~\cite{Oliver2019,kjaergaard2020}. It is commonly understood that dielectric loss in insulation layers and tunnel junction barriers contributes strongly to energy relaxation~\cite{Martinis:PRL:2005}. Even without deposited dielectrics, surface oxides and contamination at electrode interfaces are major limiting factors for qubit coherence~\cite{Quintana:APL:2014,dunsworth2017,deGraaf17,Bilmes20}.\\

Qubits require submicrometer-sized Josephson junctions to enhance the circuit's non-linearity, and to minimize the amount of lossy dielectric in the junction's tunnel barriers~\cite{Martinis:PRL:2005,Steffen2006}. Usually, such junctions are made using electron-beam patterning of photoresist to form a Dolan-bridge~\cite{Niemeyer1976,Dolan1977}, an intersection of narrow trenches~\cite{ManhattanJJ, Wu2017, kreikebaum2020}, or asymmetric undercuts~\cite{lecocq2011}. The shadow cast by these structures when metal is evaporated from two different angles then defines the junction area. After the junction's bottom electrode has been deposited, it is oxidized to form the tunnel barrier, and capped by the top electrode in the second evaporation step. Typically, submicrometer-sized junctions are fabricated on top of pre-patterned larger circuit structures such as shunt capacitors, which are made with faster UV-optical lithography~\cite{Barends13}.\\

Since the electrodes of the qubit's shunt capacitor are also oxidized during tunnel barrier formation, they are connected to the junction's top electrode through unwanted additional, so-called 'stray' junctions. 
The contribution of stray junctions to the qubit Hamiltonian is made negligible when they are much larger than the qubit junctions~\cite{Stehli2020}. Increasing the area of the stray junction also reduces the ac-voltage drop across them and thus limits dielectric loss due to structural defects in their tunnel barriers~\cite{muller2019}. Nevertheless, even large stray junctions may still contribute significantly to decoherence~\cite{Lisenfeld19}.\\

Improved qubit coherence is obtained when stray junctions are shorted using so-called bandages that are deposited in a successive lithography step~\cite{dunsworth2017}.
However, this requires additional lithography which consumes time and carries the risk of introducing further contamination.\\

Here, we describe an improved shadow-evaporation technique to fabricate sub-micrometer-sized Dolan bridge Josephson junctions together with bandage layers in a single lithography step by using three-angle evaporation. After junction formation, an argon milling plasma~\cite{Gruenhaupt2017} is applied in-situ prior to bandage deposition. Importantly, due to the anisotropy of the remote argon plasma, the sub-micrometer Josephson junction is protected in the shadow of the resist mask from damage due to impacting ions. In addition, we observe that junction aging, i.e. the drift of normal-state resistance or superconducting critical current, is significantly reduced when the junction electrodes and the bandage layers are oxidized in a controlled atmosphere directly after their deposition.\\

We note that a similar bandaging technique has recently been developed independently by Osman \emph{et al.} and demonstrated for Manhattan-style Josephson junctions~\cite{Bylander21}.\\

\begin{figure*}[htbp]
	\begin{center}
		\includegraphics[width=\textwidth]{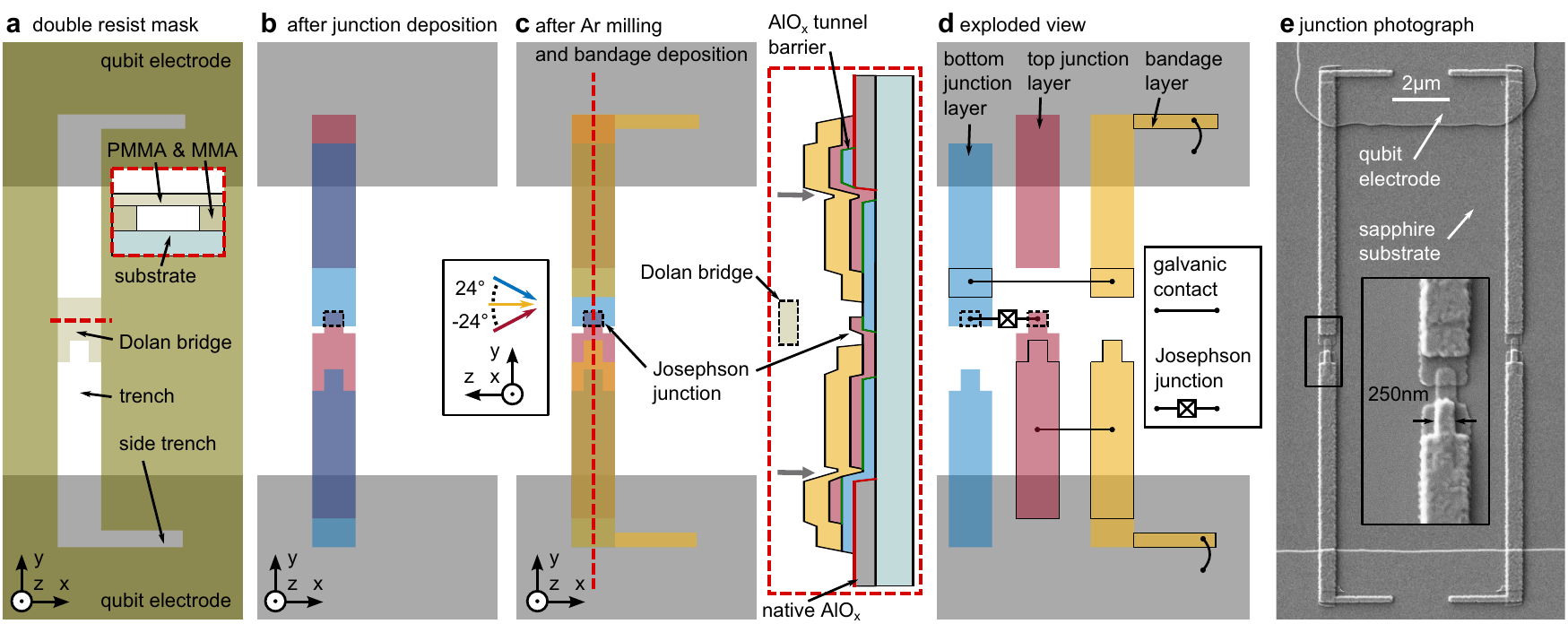}
	\end{center}
	\caption{
		\textbf{a} Sketch of the double resist mask and the Dolan bridge (see inset) for the fabrication of in-situ bandaged Josephson junctions. For clarity, the resist mask is not shown in the following sketches. \textbf{b} The Josephson junction (dashed rectangle) is created after deposition of a first (red) and second (blue) Al layers at tilted angles (see inset) and an intermediate oxidation which forms the tunnel barrier. Note that no material is deposited in the bottom of the side-trenches visible in \textbf{a}. \textbf{c} Without breaking the vacuum, the aluminum oxide is cleaned from the qubit electrodes (gray) and the junction films by argon ion milling, and a bandage film is deposited perpendicularly to the wafer. The inset shows a cross-section of the junction layers (not to scale) along the red dashed line. \textbf{d} The exploded view reveals the chain of galvanic contacts which interconnects the qubit electrodes through the Josephson contact. \textbf{e} Electron-microscopy image of a DC-SQUID consisting of two in-situ bandaged Josephson junctions, connected to the electrodes of a transmon qubit.
	}
	\label{fig:JJpic}
\end{figure*}

Figure~\ref{fig:JJpic}\textbf{a} shows the double resist mask used to form in-situ bandaged Josephson junctions (ISBJ). The Dolan bridge highlighted in green defines the junction area, underneath of which the two Al-electrodes evaporated from  $\pm24\,\mathrm{^\circ}$ angles are overlapping, see Fig.~\ref{fig:JJpic}\textbf{b}. Before the second deposition, the bottom electrode is oxidized in a static pressure of 15 mBar for 180s to form the tunnel barrier.
Note that the narrow side trenches are oriented perpendicular to the evaporation direction of the junction electrodes, so that the underlying qubit electrodes (gray) are not exposed as shown in Fig.~\ref{fig:JJpic}\textbf{b} where the mask has been omitted for clarity.\\

Next, ion-milling is applied in the same vacuum chamber to sputter the oxide from the Al films. Finally, the bandage is deposited (yellow in Fig.~\ref{fig:JJpic}\textbf{c}) perpendicularly to the substrate to create a galvanic contact between junction layers and the qubit electrodes through the side trenches. The contact areas and the Josephson contact are indicated in the legend of the exploded view shown in the right panel of Fig.~\ref{fig:JJpic}\textbf{c}. The Dolan bridge protects the junction from the argon milling and from being shorted by the bandage. An image of a dc-SQUID fabricated on sapphire, which consists\del{ing} of two ISBJs, is shown in Fig.~\ref{fig:JJpic}\textbf{e}.\\

Note that the bandage spans possible discontinuities of the junction electrodes at the film edges of the qubit electrodes, which are marked in the inset of Fig.~\ref{fig:JJpic}\textbf{c} by gray arrows. This simplifies the electrode geometry since no basewire hooks~\cite{dunsworth2017} are required which are commonly used to avoid film interruptions.\\

The critical current density of the junctions is calibrated from measurements of their normal resistance $R_\text{n}$ at room temperature via the
empirical relation $E_\text{t}=E_\text{m}(R_\text{t}/R_\text{n})^{2.5}$~\cite{Kleinsasser1995} where $E_\text{m}$ is the static oxidation exposure (mBar$\cdot$s) during tunnel barrier formation, $R_\text{t}$ is the target normal resistance, and $E_\text{t}$ the adjusted oxygen exposure.
However, the junction resistance typically shows temporal drift, known as junction aging~\cite{koppinen2007}. To explain aging, it was suggested that the tunnel barrier might incorporate aluminum hydrates~\cite{gates1984}, where the OH$^-$-group may stem from organic resist residuals~\cite{pop2012} or from water dissociation at the aluminum oxide interface~\cite{deGraaf17}.
It has also been shown that better long-term stability is obtained when junctions are annealed at a few hundred $^\circ$C temperature in vacuum~\cite{koppinen2007}, which was explained by dissociation of aluminum hydrates~\cite{gates1984}.\\

\begin{figure*}[htb]
	\begin{center}
		\includegraphics[width=.9\textwidth]{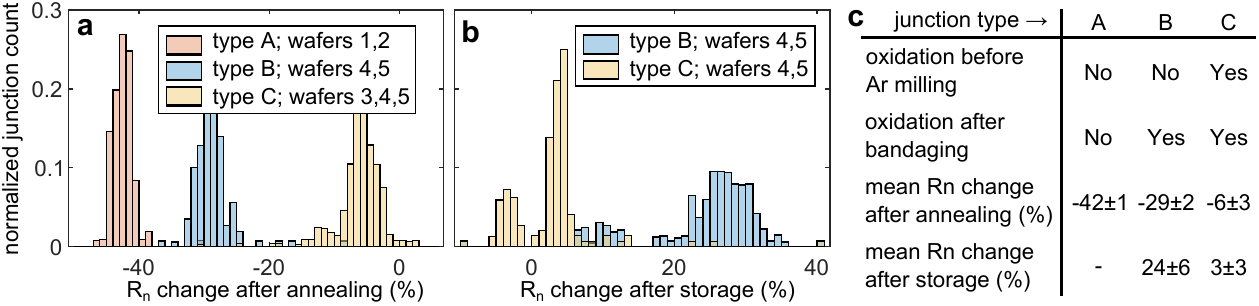}
	\end{center}
	\caption{
		\textbf{a} Resistance change of in-situ bandaged Josephson junctions after thermal annealing ($10\,\mathrm{min}$ at $200\,^\circ\mathrm{C}$ in air). Type B junctions were oxidized in-situ after bandaging,  while type C junctions were additionally oxidized after junction deposition ($10\,\mathrm{min}$ at $30\,\mathrm{mBar}$ O$_2$). Wafers 4 and 5 were cleaved in two after mask development. \textbf{b} Resistance change after storage for three weeks at ambient conditions. Type A junctions were not tested. \textbf{c} Overview of oxidation steps applied to each junction type, and their resistance change after annealing and storage.
	}
	\label{fig:RnHisto}
\end{figure*}

We monitored the stability of 500 in-situ bandaged junctions (yield~$>96\,\%$) by measuring their normal resistance $R_n$ directly after fabrication and after annealing for $10\,\mathrm{min}$ at $200\,^\circ\mathrm{C}$ in air. Junctions fabricated as described above (type A junctions) showed a resistance drop of $(42\pm2)\,\%$ after annealing. This might be due to contamination of the junction barrier by resist re-deposition during argon ion milling, and during lift-off in liquid stripper.\\

The resistance drop after junction annealing was reduced to $(29\pm2)\,\%$ when the junctions were additionally oxidized \emph{in-situ} after bandaging (type B junctions). Strongest improvement was observed when an additional oxidation step was applied between junction formation and subsequent argon ion milling ("protective oxidation", type C junctions), which reduced the resistance drop to $(6\pm3)\,\%$. We speculate that the thicker aluminum oxide may hinder contaminants from diffusing towards the tunnel barrier. Figure~\ref{fig:RnHisto}\textbf{a} summarizes the statistics of our observations for the three junction types.\\

The protective oxidation of junction electrodes also improves the long-term stability during storage. After three weeks at ambient conditions (stored in a laboratory drawer), the resistance of type B junctions increased by $(24\pm6)\,\%$, while type C junctions showed an increase of $(3\pm3)\,\%$ (see Fig.~\ref{fig:RnHisto}\textbf{b}).\\

\begin{figure}[htb]
	\begin{center}
		\includegraphics[width=.38\textwidth]{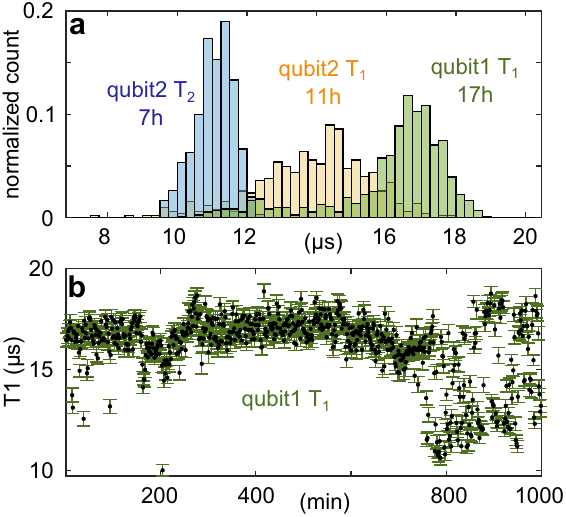}
	\end{center}
	\caption{
		\textbf{a} Histograms of energy relaxation time $T_1$ and Ramsey dephasing time $T_2$ recorded over several hours (the duration is indicated in each histogram label), with two out of three qubits from a same qubit chip. Qubit three was tuned far away from the Transmon sweet spot due to trapped vortices. \textbf{b} $T_1$ records vs. time which result in the histogram for qubit No. 1 in the panel above. We recognize that the qubit experienced strong coherence fluctuations in the last three hours of records, most probably due to coupling to some charged tunneling defects~\cite{Klimov2018,Schloer2019}.
	}
	\label{fig:T1Histo}
\end{figure}

To test the suitability of in-situ bandaged Josephson junctions for quantum bits, we used them to fabricate Xmon-type transmon qubits~\cite{Barends13,KochTransmon}. These had charge and Josephson energies designed to $E_\text{C}\sim200\,\mathrm{MHz}$ and $E_\text{J}\sim20\,\mathrm{GHz}$, respectively.
We monitored their energy relaxation times $T_1$ at qubit resonance frequencies of about $6\ \mathrm{GHz}$ during several hours to account for temporal fluctuations due to the interference with material defects~\cite{Lisenfeld_2019,burnett2019}.
Figure~\ref{fig:T1Histo} \textbf{a} shows histograms of $T_1$ for two tested samples. The obtained average $T_1$ times were very similar than those we obtained on similarly designed qubits~\cite{bilmes2021probing} that were fabricated either with classical shadow junctions~\cite{Niemeyer1976,Dolan1977} or cross-type junctions~\cite{Steffen2005,Wu2017}. Thus, ISBJs did not generate identifiable excess dielectric loss. Figure.~\ref{fig:T1Histo} \textbf{b} contains the time evolution of $T_1$ of qubit No. 1, showing strong fluctuations during the last hours of measurement. This is a common issue that is explained by resonance frequency fluctuations of strongly interacting tunneling  defects residing on qubit electrodes~\cite{Klimov2018,Schloer2019}.\\

We conclude that the presented technique is applicable for fabrication of coherent qubits that are free of stray Josephson junctions, and it works reliably and economizes one lithography step. The reported method also preserves the advantage of conventional bandaging~\cite{dunsworth2017} where the interface of junction films to the substrate was not harmed by argon ion milling, and it is slightly simpler than the in-situ bandaging technique~\cite{Bylander21} based on Manhattan-style junctions~\cite{ManhattanJJ}, as it requires a uniaxial wafer tilt for shadow evaporation. Moreover, we observed that the changes of junction resistance induced by thermal annealing and by long-term storage are significantly reduced by adding oxidation steps after junction and bandage layer depositions.\\

These results offer improvements in the fabrication of stable and contamination-free Josephson junctions as required by quantum-limited parametric amplifiers~\cite{sweeny1985,macklin2015} and superconducting quantum processors. The in-situ bandaging technique to avoid parasitic tunnel barriers can also facilitate the deposition of multi-material stacks, e.g. to fabricate superconductor-ferromagnet junctions which have applications in spintronics~\cite{linder2015} and superconducting logic circuits~\cite{feofanov2010}. It is also suitable for so-called cross junctions whose bottom layer is deposited separately~\cite{Steffen2006,Wu2017}, where the top and bandaging layers can be deposited under distinct wafer orientations.\\

\subsection*{Author Contributions}
The fabrication method and design for ISBJs was developed by AB, and fabricated and tested by AB, SV and AKN. Measurements on qubits were done by JL. The manuscript was written by AB and JL with contributions from all authors.

\subsection*{Acknowledgements}
The authors gratefully acknowledge funding by Google LLC, and support by the KIT-Publication Fund of the Karlsruhe Institute of Technology. AVU acknowledges partial support provided by Rosatom, and the Ministry of Education and Science of the Russian Federation in the framework of the Program to Increase Competitiveness of the NUST MISIS (contract No. K2-2020-022).

\subsection*{Supplementary Material}
See supplementary material for detailed fabrication recipe of the studied in-situ bandaged Josephson junctions.

\subsection*{Data Availability}
The data that support the findings of this study are available from the corresponding author upon reasonable request.

\bibliography{Biblio}

\begin{thebibliography}{10}
\expandafter\ifx\csname url\endcsname\relax
  \def\url#1{\texttt{#1}}\fi
\expandafter\ifx\csname urlprefix\endcsname\relax\def\urlprefix{URL }\fi
\providecommand{\bibinfo}[2]{#2}
\providecommand{\eprint}[2][]{\url{#2}}

\bibitem{Oliver2019}
\bibinfo{author}{Krantz, P.} \emph{et~al.}
\newblock \bibinfo{title}{A quantum engineer's guide to superconducting
  qubits}.
\newblock \emph{\bibinfo{journal}{Applied Physics Reviews}}
  \textbf{\bibinfo{volume}{6}}, \bibinfo{pages}{021318} (\bibinfo{year}{2019}).
\newblock \urlprefix\url{https://doi.org/10.1063/1.5089550}.
\newblock \eprint{https://doi.org/10.1063/1.5089550}.

\bibitem{kjaergaard2020}
\bibinfo{author}{Kjaergaard, M.} \emph{et~al.}
\newblock \bibinfo{title}{Superconducting qubits: Current state of play}.
\newblock \emph{\bibinfo{journal}{Annual Review of Condensed Matter Physics}}
  \textbf{\bibinfo{volume}{11}}, \bibinfo{pages}{369--395}
  (\bibinfo{year}{2020}).

\bibitem{Martinis:PRL:2005}
\bibinfo{author}{Martinis, J.~M.} \emph{et~al.}
\newblock \bibinfo{title}{{Decoherence in Josephson qubits from dielectric
  loss}}.
\newblock \emph{\bibinfo{journal}{Physical Review Letters}}
  \textbf{\bibinfo{volume}{95}}, \bibinfo{pages}{210503}
  (\bibinfo{year}{2005}).

\bibitem{Quintana:APL:2014}
\bibinfo{author}{Quintana, C.~M.} \emph{et~al.}
\newblock \bibinfo{title}{{Characterization and reduction of
  microfabrication-induced decoherence in superconducting quantum circuits}}.
\newblock \emph{\bibinfo{journal}{Appl. Phys. Lett.}}
  \textbf{\bibinfo{volume}{105}}, \bibinfo{pages}{062601}
  (\bibinfo{year}{2014}).
\newblock
  \urlprefix\url{http://scitation.aip.org/content/aip/journal/apl/105/6/10.1063/1.4893297}.

\bibitem{dunsworth2017}
\bibinfo{author}{Dunsworth, A.} \emph{et~al.}
\newblock \bibinfo{title}{Characterization and reduction of capacitive loss
  induced by sub-micron josephson junction fabrication in superconducting
  qubits}.
\newblock \emph{\bibinfo{journal}{Applied Physics Letters}}
  \textbf{\bibinfo{volume}{111}}, \bibinfo{pages}{022601}
  (\bibinfo{year}{2017}).

\bibitem{deGraaf17}
\bibinfo{author}{de~Graaf, S.~E.} \emph{et~al.}
\newblock \bibinfo{title}{Direct identification of dilute surface spins on
  ${\mathrm{al}}_{2}{\mathrm{o}}_{3}$: Origin of flux noise in quantum
  circuits}.
\newblock \emph{\bibinfo{journal}{Phys. Rev. Lett.}}
  \textbf{\bibinfo{volume}{118}}, \bibinfo{pages}{057703}
  (\bibinfo{year}{2017}).
\newblock
  \urlprefix\url{https://link.aps.org/doi/10.1103/PhysRevLett.118.057703}.

\bibitem{Bilmes20}
\bibinfo{author}{Bilmes, A.} \emph{et~al.}
\newblock \bibinfo{title}{Resolving the positions of defects in superconducting
  quantum bits}.
\newblock \emph{\bibinfo{journal}{Scientific Reports}}
  \textbf{\bibinfo{volume}{10}}, \bibinfo{pages}{1--6} (\bibinfo{year}{2020}).

\bibitem{Steffen2006}
\bibinfo{author}{Steffen, M.} \emph{et~al.}
\newblock \bibinfo{title}{State tomography of capacitively shunted phase qubits
  with high fidelity}.
\newblock \emph{\bibinfo{journal}{Phys. Rev. Lett.}}
  \textbf{\bibinfo{volume}{97}}, \bibinfo{pages}{050502}
  (\bibinfo{year}{2006}).
\newblock
  \urlprefix\url{https://link.aps.org/doi/10.1103/PhysRevLett.97.050502}.

\bibitem{Niemeyer1976}
\bibinfo{author}{Niemeyer, J.} \& \bibinfo{author}{Kose, V.}
\newblock \bibinfo{title}{Observation of large dc supercurrents at nonzero
  voltages in josephson tunnel junctions}.
\newblock \emph{\bibinfo{journal}{Applied Physics Letters}}
  \textbf{\bibinfo{volume}{29}}, \bibinfo{pages}{380--382}
  (\bibinfo{year}{1976}).
\newblock \urlprefix\url{https://doi.org/10.1063/1.89094}.
\newblock \eprint{https://doi.org/10.1063/1.89094}.

\bibitem{Dolan1977}
\bibinfo{author}{Dolan, G.~J.}
\newblock \bibinfo{title}{Offset masks for lift-off photoprocessing}.
\newblock \emph{\bibinfo{journal}{Applied Physics Letters}}
  \textbf{\bibinfo{volume}{31}}, \bibinfo{pages}{337--339}
  (\bibinfo{year}{1977}).
\newblock \urlprefix\url{https://doi.org/10.1063/1.89690}.
\newblock \eprint{https://doi.org/10.1063/1.89690}.

\bibitem{ManhattanJJ}
\bibinfo{author}{Potts, A.}, \bibinfo{author}{Parker, G.},
  \bibinfo{author}{Baumberg, J.} \& \bibinfo{author}{de~Groot, P.}
\newblock \bibinfo{title}{Cmos compatible fabrication methods for submicron
  josephson junction qubits}.
\newblock \emph{\bibinfo{journal}{IEE Proceedings-Science, Measurement and
  Technology}} \textbf{\bibinfo{volume}{148}}, \bibinfo{pages}{225--228}
  (\bibinfo{year}{2001}).

\bibitem{Wu2017}
\bibinfo{author}{Wu, X.} \emph{et~al.}
\newblock \bibinfo{title}{Overlap junctions for high coherence superconducting
  qubits}.
\newblock \emph{\bibinfo{journal}{Applied Physics Letters}}
  \textbf{\bibinfo{volume}{111}}, \bibinfo{pages}{032602}
  (\bibinfo{year}{2017}).

\bibitem{kreikebaum2020}
\bibinfo{author}{Kreikebaum, J.}, \bibinfo{author}{O’Brien, K.},
  \bibinfo{author}{Morvan, A.} \& \bibinfo{author}{Siddiqi, I.}
\newblock \bibinfo{title}{Improving wafer-scale josephson junction resistance
  variation in superconducting quantum coherent circuits}.
\newblock \emph{\bibinfo{journal}{Superconductor Science and Technology}}
  \textbf{\bibinfo{volume}{33}}, \bibinfo{pages}{06LT02}
  (\bibinfo{year}{2020}).

\bibitem{lecocq2011}
\bibinfo{author}{Lecocq, F.} \emph{et~al.}
\newblock \bibinfo{title}{Junction fabrication by shadow evaporation without a
  suspended bridge}.
\newblock \emph{\bibinfo{journal}{Nanotechnology}}
  \textbf{\bibinfo{volume}{22}}, \bibinfo{pages}{315302}
  (\bibinfo{year}{2011}).

\bibitem{Barends13}
\bibinfo{author}{Barends, R.} \emph{et~al.}
\newblock \bibinfo{title}{Coherent josephson qubit suitable for scalable
  quantum integrated circuits}.
\newblock \emph{\bibinfo{journal}{Phys. Rev. Lett.}}
  \textbf{\bibinfo{volume}{111}}, \bibinfo{pages}{080502}
  (\bibinfo{year}{2013}).

\bibitem{Stehli2020}
\bibinfo{author}{Stehli, A.} \emph{et~al.}
\newblock \bibinfo{title}{Coherent superconducting qubits from a subtractive
  junction fabrication process}.
\newblock \emph{\bibinfo{journal}{Applied Physics Letters}}
  \textbf{\bibinfo{volume}{117}}, \bibinfo{pages}{124005}
  (\bibinfo{year}{2020}).
\newblock \urlprefix\url{https://doi.org/10.1063/5.0023533}.
\newblock \eprint{https://doi.org/10.1063/5.0023533}.

\bibitem{muller2019}
\bibinfo{author}{M{\"u}ller, C.}, \bibinfo{author}{Cole, J.~H.} \&
  \bibinfo{author}{Lisenfeld, J.}
\newblock \bibinfo{title}{Towards understanding two-level-systems in amorphous
  solids: insights from quantum circuits}.
\newblock \emph{\bibinfo{journal}{Reports on Progress in Physics}}
  \textbf{\bibinfo{volume}{82}}, \bibinfo{pages}{124501}
  (\bibinfo{year}{2019}).

\bibitem{Lisenfeld19}
\bibinfo{author}{Lisenfeld, J.} \emph{et~al.}
\newblock \bibinfo{title}{Electric field spectroscopy of material defects in
  transmon qubits}.
\newblock \emph{\bibinfo{journal}{npj Quantum Information}}
  \textbf{\bibinfo{volume}{5}}, \bibinfo{pages}{1--6} (\bibinfo{year}{2019}).

\bibitem{Gruenhaupt2017}
\bibinfo{author}{Gr\"unhaupt, L.} \emph{et~al.}
\newblock \bibinfo{title}{An argon ion beam milling process for native alox
  layers enabling coherent superconducting contacts}.
\newblock \emph{\bibinfo{journal}{Applied Physics Letters}}
  \textbf{\bibinfo{volume}{111}}, \bibinfo{pages}{072601}
  (\bibinfo{year}{2017}).
\newblock \urlprefix\url{https://doi.org/10.1063/1.4990491}.
\newblock \eprint{https://doi.org/10.1063/1.4990491}.

\bibitem{Bylander21}
\bibinfo{author}{Osman, A.} \emph{et~al.}
\newblock \bibinfo{title}{Simplified josephson-junction fabrication process for
  reproducibly high-performance superconducting qubits}.
\newblock \emph{\bibinfo{journal}{Applied Physics Letters}}
  \textbf{\bibinfo{volume}{118}}, \bibinfo{pages}{064002}
  (\bibinfo{year}{2021}).
\newblock \urlprefix\url{https://doi.org/10.1063/5.0037093}.
\newblock \eprint{https://doi.org/10.1063/5.0037093}.

\bibitem{Kleinsasser1995}
\bibinfo{author}{Kleinsasser, A.~W.}, \bibinfo{author}{Miller, R.~E.} \&
  \bibinfo{author}{Mallison, W.~H.}
\newblock \bibinfo{title}{Dependence of critical current density on oxygen
  exposure in nb-alo/sub x/-nb tunnel junctions}.
\newblock \emph{\bibinfo{journal}{IEEE transactions on applied
  superconductivity}} \textbf{\bibinfo{volume}{5}}, \bibinfo{pages}{26--30}
  (\bibinfo{year}{1995}).

\bibitem{koppinen2007}
\bibinfo{author}{Koppinen, P.}, \bibinfo{author}{V{\"a}ist{\"o}, L.} \&
  \bibinfo{author}{Maasilta, I.}
\newblock \bibinfo{title}{Complete stabilization and improvement of the
  characteristics of tunnel junctions by thermal annealing}.
\newblock \emph{\bibinfo{journal}{Applied physics letters}}
  \textbf{\bibinfo{volume}{90}}, \bibinfo{pages}{053503}
  (\bibinfo{year}{2007}).

\bibitem{gates1984}
\bibinfo{author}{Gates, J.}, \bibinfo{author}{Washington, M.} \&
  \bibinfo{author}{Gurvitch, M.}
\newblock \bibinfo{title}{Critical current uniformity and stability of
  nb/al-oxide-nb josephson junctions}.
\newblock \emph{\bibinfo{journal}{Journal of applied physics}}
  \textbf{\bibinfo{volume}{55}}, \bibinfo{pages}{1419--1421}
  (\bibinfo{year}{1984}).

\bibitem{pop2012}
\bibinfo{author}{Pop, I.-M.} \emph{et~al.}
\newblock \bibinfo{title}{Fabrication of stable and reproducible submicron
  tunnel junctions}.
\newblock \emph{\bibinfo{journal}{Journal of Vacuum Science \& Technology B,
  Nanotechnology and Microelectronics: Materials, Processing, Measurement, and
  Phenomena}} \textbf{\bibinfo{volume}{30}}, \bibinfo{pages}{010607}
  (\bibinfo{year}{2012}).

\bibitem{Klimov2018}
\bibinfo{author}{Klimov, P.~V.} \emph{et~al.}
\newblock \bibinfo{title}{Fluctuations of energy-relaxation times in
  superconducting qubits}.
\newblock \emph{\bibinfo{journal}{Phys. Rev. Lett.}}
  \textbf{\bibinfo{volume}{121}}, \bibinfo{pages}{090502}
  (\bibinfo{year}{2018}).
\newblock
  \urlprefix\url{https://link.aps.org/doi/10.1103/PhysRevLett.121.090502}.

\bibitem{Schloer2019}
\bibinfo{author}{Schl\"or, S.} \emph{et~al.}
\newblock \bibinfo{title}{Correlating decoherence in transmon qubits: Low
  frequency noise by single fluctuators}.
\newblock \emph{\bibinfo{journal}{Phys. Rev. Lett.}}
  \textbf{\bibinfo{volume}{123}}, \bibinfo{pages}{190502}
  (\bibinfo{year}{2019}).
\newblock
  \urlprefix\url{https://link.aps.org/doi/10.1103/PhysRevLett.123.190502}.

\bibitem{KochTransmon}
\bibinfo{author}{Koch, J.} \emph{et~al.}
\newblock \bibinfo{title}{Charge-insensitive qubit design derived from the
  cooper pair box}.
\newblock \emph{\bibinfo{journal}{Physical Review A}}
  \textbf{\bibinfo{volume}{76}}, \bibinfo{pages}{042319}
  (\bibinfo{year}{2007}).

\bibitem{Lisenfeld_2019}
\bibinfo{author}{Müller, C.}, \bibinfo{author}{Cole, J.~H.} \&
  \bibinfo{author}{Lisenfeld, J.}
\newblock \bibinfo{title}{Towards understanding two-level-systems in amorphous
  solids: insights from quantum circuits}.
\newblock \emph{\bibinfo{journal}{Reports on Progress in Physics}}
  \textbf{\bibinfo{volume}{82}}, \bibinfo{pages}{124501}
  (\bibinfo{year}{2019}).
\newblock \urlprefix\url{https://doi.org/10.1088/1361-6633/ab3a7e}.

\bibitem{burnett2019}
\bibinfo{author}{Burnett, J.~J.} \emph{et~al.}
\newblock \bibinfo{title}{Decoherence benchmarking of superconducting qubits}.
\newblock \emph{\bibinfo{journal}{npj Quantum Information}}
  \textbf{\bibinfo{volume}{5}}, \bibinfo{pages}{54} (\bibinfo{year}{2019}).
\newblock \urlprefix\url{https://doi.org/10.1038/s41534-019-0168-5}.

\bibitem{bilmes2021probing}
\bibinfo{author}{Bilmes, A.}, \bibinfo{author}{Volosheniuk, S.},
  \bibinfo{author}{Ustinov, A.~V.} \& \bibinfo{author}{Lisenfeld, J.}
\newblock \bibinfo{title}{Probing defect densities at the edges and inside
  josephson junctions of superconducting qubits} (\bibinfo{year}{2021}).
\newblock \eprint{2108.06555}.

\bibitem{Steffen2005}
\bibinfo{author}{Steffen, M.} \emph{et~al.}
\newblock \bibinfo{title}{State tomography of capacitively shunted phase qubits
  with high fidelity}.
\newblock \emph{\bibinfo{journal}{Phys. Rev. Lett.}}
  \textbf{\bibinfo{volume}{97}}, \bibinfo{pages}{050502}
  (\bibinfo{year}{2006}).
\newblock
  \urlprefix\url{http://link.aps.org/doi/10.1103/PhysRevLett.97.050502}.

\bibitem{sweeny1985}
\bibinfo{author}{Sweeny, M.} \& \bibinfo{author}{Mahler, R.}
\newblock \bibinfo{title}{A travelling-wave parametric amplifier utilizing
  josephson junctions}.
\newblock \emph{\bibinfo{journal}{IEEE Transactions on Magnetics}}
  \textbf{\bibinfo{volume}{21}}, \bibinfo{pages}{654--655}
  (\bibinfo{year}{1985}).

\bibitem{macklin2015}
\bibinfo{author}{Macklin, C.} \emph{et~al.}
\newblock \bibinfo{title}{A near--quantum-limited josephson traveling-wave
  parametric amplifier}.
\newblock \emph{\bibinfo{journal}{Science}} \textbf{\bibinfo{volume}{350}},
  \bibinfo{pages}{307--310} (\bibinfo{year}{2015}).

\bibitem{linder2015}
\bibinfo{author}{Linder, J.} \& \bibinfo{author}{Robinson, J.~W.}
\newblock \bibinfo{title}{Superconducting spintronics}.
\newblock \emph{\bibinfo{journal}{Nature Physics}}
  \textbf{\bibinfo{volume}{11}}, \bibinfo{pages}{307--315}
  (\bibinfo{year}{2015}).

\bibitem{feofanov2010}
\bibinfo{author}{Feofanov, A.} \emph{et~al.}
\newblock \bibinfo{title}{Implementation of
  superconductor/ferromagnet/superconductor $\pi$-shifters in superconducting
  digital and quantum circuits}.
\newblock \emph{\bibinfo{journal}{Nature Physics}}
  \textbf{\bibinfo{volume}{6}}, \bibinfo{pages}{593--597}
  (\bibinfo{year}{2010}).

\bibitem{Rooks2002}
\bibinfo{author}{Rooks, M.~J.} \emph{et~al.}
\newblock \bibinfo{title}{Low stress development of poly(methylmethacrylate)
  for high aspect ratio structures}.
\newblock \emph{\bibinfo{journal}{Journal of Vacuum Science \& Technology B:
  Microelectronics and Nanometer Structures Processing, Measurement, and
  Phenomena}} \textbf{\bibinfo{volume}{20}}, \bibinfo{pages}{2937--2941}
  (\bibinfo{year}{2002}).
\newblock \urlprefix\url{https://avs.scitation.org/doi/abs/10.1116/1.1524971}.
\newblock \eprint{https://avs.scitation.org/doi/pdf/10.1116/1.1524971}.

\end{thebibliography}

\clearpage
\onecolumngrid
\renewcommand{\thefigure}{\textbf{S\arabic{figure}}}
\renewcommand{\thesection}{\textbf{\Alph{section}}}
\setcounter{figure}{0}

\begin{center}
	\large
	\textbf{In-situ bandaged Josephson junctions for superconducting quantum processors\\}
	\vspace{0.3cm}
	\textbf{Supplementary Material}
	\normalsize
\end{center}

\section{Sample fabrication}
\label{ss_fab}
\textbf{Wafer preparation and optical lithography}\\
The 3", C-plane, DSP, 500um sapphire wafer was washed in piranha solution and then cleaned in a barrel asher (10 min, 45 sccm O$_2$, 150 W, 0.7 mbar). Loaded into the vacuum chamber of a PLASSYS evaporation tool, the wafer was baked for 2 hours at $200\,\mathrm{^\circ C}$, then cooled down over night ($\sim$ 12h). Further, a gentle O$_2$/Ar cleaning plasma was applied for $10\,\mathrm{s}$ with a Kaufmann source
, and after successive Titanium gettering, 100nm of aluminum were evaporated (at $1\,\mathrm{nm/s}$) to form the ground plane.\\

The large structures (transmission line, resonators, qubit electrodes) with a critical dimension of $2.5\,\mathrm{\mu m}$ were patterned into the Al film with an Ar-Cl inductively coupled plasma, using an S1805 resist mask which was prepared in an optical lithography step.\\

\textbf{Formation of the Josephson junctions}\\
An MMA/PMMA double resist mask with a critical dimension of $250\,\mathrm{nm}$ was prepared in an electron-beam lithography. After development of the mask, the PMMA residuals were removed in the barrel asher (3 min, 8 sccm O$_2$, 270 W, 0.6 mbar), after which the wafer was loaded into the PLASSYS and pumped to a loadlock pressure of ~3e-7mbar. After a short plasma cleaning and Titanium gettering as described above, the ISBJ is formed. The film thicknesses of the first and second junction electrodes were $30\,\mathrm{nm}$ each while the bandage was $140\,\mathrm{nm}$ thick.\\

\textbf{Detailed fabrication steps}\\
Three junction types (A,B and C) were studied which differ by extra oxidation steps before and after bandaging, as described below.\\
\begin{enumerate}
	\item Wafer preparation (3"):
	\begin{enumerate}
		\item Clean with piranha solution (mix of sulfuric acid H$_2$SO$_4$ and hydrogen peroxide H$_2$O$_2$) to remove organic residuals.
		\item Oxygen plasma clean in the "barrel asher" (see Tab~\ref{tab_fab}): $10\,\minute$ duration, $45\,\mathrm{sccm}$ O$_2$ at a chamber pressure of $0.7\,\milli\mathrm{Bar}$, $150\,\watt$ generator power.
		\item After the O2 clean, transfer ($\sim 5\,\minute$) the wafer into the PLASSYS shadow evaporator and pump the load lock for $2\,\hour$ at $200\,\mathrm{^\circ C}$. Cool down to room temperature over night. Final pressure is $\sim1\cdot10^{-7}\,\milli\mathrm{Bar}$.
		\item Apply oxygen-argon cleaning recipe (Table~\ref{tab_descum}) for $10\,\mathrm{sec}$ at $45^\circ$ tilt angle.
		\item Ti gettering (deposit $\sim30\,\nano\meter$ Ti into the load lock with closed shutter, and wait for $5\,\minute$).
		\item Deposit $\sim100\,\nano\meter$ Al that will define the qubit ground plane (zero tilt, deposition rate of $1\,\nano\meter/\second$).
		\item Apply static oxidation ($10\,\minute$ at $30\,\milli\mathrm{Bar}$) to passivate the Al film surface.
		\item Store the wafer at ambient conditions for at least one hour to stabilize the native oxide of the Al film.
		\item Spin coat the wafer with S1818 protecting resist, and dice it into seven $2\,\mathrm{cm}\times2\,\mathrm{cm}$ wafers.\label{itemDice}
		\item Remove the protecting resist in a NEP bath ($1\,\hour$ at $90\,^\circ\mathrm{C}$).
	\end{enumerate}
	\item Definition of qubit electrodes and readout resonator:
	\begin{enumerate}
		\item Spin coat the wafer ($2\,\mathrm{cm}\times2\,\mathrm{cm}$) with S1805 resist ($\sim 300\,\nano\meter$ thick, $6$k rpm, $60$ sec, 1 min on hotplate at 115 $^\circ$C), and apply positive optical lithography using the mask aligner (Tab.~\ref{tab_fab}).
		\item Etch the Al film using an Ar-Cl plasma in an ICP device (Tab.~\ref{tab_fab}).
		\item Remove the resist in a NEP bath ($1\,\hour$ at $90\,^\circ\mathrm{C}$).
	\end{enumerate}
	\item Deposition of the in-situ bandaged Josephson junctions:
	\begin{enumerate}
		\item Spin coat the wafer with a double resist ($\sim250\,\nano\meter$ PMMA A-4 on top of $\sim900\,\nano\meter$ MMA EL-13, each spun at $2$k rpm for $100$ sec, and baked for $5$ min at $200\,^\circ$C), and cover with a thin Gold film to improve electric conductivity in the following positive electron-beam lithography step (JEOL device, see Tab.~\ref{tab_fab}). Develop the mask in a cold Isopropanol/water mixture (3/2) at $6^\circ$C~\cite{Rooks2002}.
		\item Oxygen plasma clean with the barrel asher ($3\,\minute$ duration, $7.5\,\mathrm{sccm}$ O$_2$, $230\,\watt$ generator power) to remove resist residuals.
		\item Load the wafer into the PLASSYS and pump the load lock for $\sim2\,\hour$ at room temperature to a pressure of $\sim4\cdot 10^{-7}\,\milli\mathrm{Bar}$.
		\item Apply oxygen-argon recipe (Table~\ref{tab_descum}) for $5\,\mathrm{sec}$ at $50^\circ$ tilt, in order to clean the wafer underneath the Dolan bridge of the resist mask.
		\item Ti gettering.
		\item Deposit $30\,\nano\meter$ Aluminum at $24^\circ$, and at a rate of $1\,\nano\meter/\second$, in order to define the junction's bottom electrode.
		\item Apply static oxidation (exposure $180\,\second$ at $15\,\milli\mathrm{Bar}$) to form the AlO$_x$ tunnel barrier.
		\item Deposit $30\,\nano\meter$ Aluminum at $-24^\circ$, and at a rate of $1\,\nano\meter/\second$, to form the junction's top electrode.
		\item For junction type C, apply a static oxidation ($10\,\minute$ at $30\,\milli\mathrm{Bar}$) to passivate the junction's surface, and protect it from contamination during in-situ bandaging.
		\item Apply argon milling (Table~\ref{tab_armill}) for $2.5\,\minute$ at zero tilt, in order to remove the oxide from the junction electrodes, and native oxide from the qubit electrodes.
		\item Deposit $140\,\nano\meter$ Aluminum at zero tilt, to define the bandage which galvanically interconnects all junction layers and the qubit electrodes.
		\item For junction type B and C, apply a static oxidation ($10\,\minute$ at $30\,\milli\mathrm{Bar}$) to passivate the junction's surface, and protect it from contamination during liftoff.
		\item Lift-off in a NEP bath ($1\,\hour$ at $90\,^\circ\mathrm{C}$).
	\end{enumerate}
	\item dicing into small chips, as described in \ref{itemDice}.
\end{enumerate}

\begin{table}[hp]
	\centering
	\begin{tabular}{|c|c|}
		\hline
		Device & Model \\ \hline\hline 
		electron beam writer & JEOL JBX-5500ZD\\
		& $50\,\mathrm{keV}$ acceleration voltage\\ \hline
		mask aligner &  Carl Suess MA6, Xe $500\,\mathrm{W}$ lamp,\\
		& wave lengths: $240,\;365,\;405\,\mathrm{nm}$\\ \hline
		inductively coupled & Oxford Plasma Technology\\
		plasma (ICP) device & Plasmalab 100 ICP180\\
		&variable substrate temp. $273-343\,\mathrm{K}$\\
		\hline 
		shadow evaporation device & PLASSYS MEB550s\\
		\hline
		O$_2$ cleaner ("Barrel asher") & "nano", Diener electronic GmbH\\
		&$40\,\kilo\mathrm{Hz}$ generator, $0-300\,\watt$\\
		\hline  
	\end{tabular}
	\caption{Apparatus used at KIT to fabricate qubit samples.}
	\label{tab_fab}
\end{table}

\begin{table}[h!]
	\centering
	\begin{tabular}{|c|c|c|c|c|}
		\hline
		\multicolumn{5}{|l|}{\textbf{Descum} ($10\,\mathrm{sccm}$ O$_2$, $5\,\mathrm{sccm}$ Ar)}\\ \hline
		Cathode & Discharge & Beam & acceleration & neutralizer \\
		$8\,\volt$ & $40\,\volt$ & $50\,\volt$ & $11\,\volt$ & $5\,\volt$ (emission) \\
		$5\,\ampere$ & $0.1\,\ampere$ & $5\,\milli\ampere$ & $4\,\milli\ampere$ & $9\,\volt$ \\ 
		& & & & $11\,\milli\ampere$ \\ \hline
	\end{tabular}
	\caption{Parameters of the KSC 1202 power supply unit that controls the KDC 40 ion source to generate the remote Oxygen-Argon plasma used to descum or clean samples in the PLASSYS MEB550s shadow evaporation tool.}
	\label{tab_descum}
\end{table}

\begin{table}[h!]
	\centering
	\begin{tabular}{|c|c|c|c|c|}
		\hline
		\multicolumn{5}{|l|}{\textbf{Argon ion-milling} ($4\,\mathrm{sccm}$ Arr)}\\ \hline
		Cathode & Discharge & Beam & acceleration & neutralizer \\
		$7\,\volt$ & $50\,\volt$ & $400\,\volt$ & $91\,\volt$ & $15\,\volt$ (emission) \\
		$4\,\ampere$ & $0.1\,\ampere$ & $15\,\milli\ampere$ & $0.9\,\milli\ampere$ & $9\,\volt$ \\ 
		& & & & $1\,\milli\ampere$ \\ \hline
	\end{tabular}
	\caption{Parameters of the KSC 1202 power supply unit that controls the KDC 40 ion source to generate the Argon ion-milling remote plasma in the PLASSYS MEB550s shadow evaporation device.}
	\label{tab_armill}
\end{table}

\end{document}